# Vers un pilotage de la qualité des pièces injectées

Pierre Nagorny
Université Savoie Mont Blanc
Laboratoire SYMME, 7 Chemin de Bellevue
Annecy-le-Vieux – France
pierre.nagorny@univ-smb.fr

Eric Pairel
Université Savoie Mont Blanc
Laboratoire SYMME, 7 Chemin de Bellevue
Annecy-le-Vieux – France
eric.pairel@univ-smb.fr

Maurice Pillet
Université Savoie Mont Blanc
Laboratoire SYMME, 7 Chemin de Bellevue
Annecy-le-Vieux – France
maurice.pillet@univ-smb.fr

*Résumé— Le procédé d'injection des thermoplastiques permet la production de pièces complexes en grandes séries. Les exigences qualités sont croissantes. Il est nécessaire de réguler le procédé d'injection afin de conserver un point de fonctionnement. Le pilotage du procédé ne permet pas, aujourd'hui, de proposer d'ajuster l'ensemble des paramètres du procédé pour optimiser la qualité du produit. Afin de proposer une méthode robuste de pilotage auto-adaptative, nous pouvons nous appuyer sur les succès de la modélisation par réseau de neurones. L'objectif est de piloter la machine cycle à cycle à partir de caractéristiques mesurées sur la pièce produite. Le temps de cycle est souvent inférieur à 30 secondes pour ce procédé de fabrication ; ce qui pose le défi du mesurage, du traitement et de l'ajustement des paramètres dans ce court délai. Cette présentation nous permet d'établir une étude de la littérature sur laquelle s'appuiera notre démarche expérimentale.*

*Mots-clés— injection des thermoplastiques, pilotage, réseau de neurones, plan d'expériences*

## I. Introduction

Le procédé d'injection des thermoplastiques permet de produire des pièces complexes en grandes séries. Il demande une qualification longue de l'outillage et le savoir-faire du technicien afin de régler les paramètres du procédé. Ces paramètres influent de manière non-linéaire sur les caractéristiques de la pièce finale. Plusieurs facteurs non contrôlables agissent comme perturbations et doivent être compensés par l'ajustement des paramètres. Garantir la qualité des pièces en sortie machine est cruciale car elle détermine la suite des opérations de fabrication et la pièce finale. Le procédé d'injection se compose de phases interdépendantes : les paramètres d'une phase influent sur la phase suivante. Dans l'ensemble de la littérature, il n'a jamais été proposé d'ajuster l'ensemble des paramètres réglables du procédé afin de garantir l'ensemble des caractéristiques qualités attendues du produit. Nous investiguons une méthode de pilotage du procédé cycle à cycle basée sur la mesure de la qualité des pièces produites. Cette approche globale peut s'appuyer sur les succès de la modélisation du procédé par réseaux de neurones, sur la faisabilité du mesurage en cycle et sur l'analyse statistique afin d'optimiser la quantité à ajuster. Une approche globale permettra de trouver un point de fonctionnement idéal du procédé, qu'il conviendra de conserver par régulation.

### A. Interdépendances des phases du procédé

Nous proposons une représentation (Figure 1) Zig Zag Process dans les deux derniers piliers définis par l'approche *Axiomatic Design* [1] qui fait apparaître les enjeux transverses du pilotage du procédé : les paramètres du procédé {Pi} et les caractéristiques des produits lors de chacune des phases {Ci}. Les caractéristiques du produit fini {C} sont fonction de l'ensemble des phases. De plus, chaque phase est fonction des phases précédentes et des variables du procédé. Si un réglage est modifié au niveau de l'une des phases, les caractéristiques du procédé pour toutes les phases avales seront modifiées. Dans la suite de notre présentation, nous indiquerons à quelle phase de notre représentation correspondent les variables des études citées.

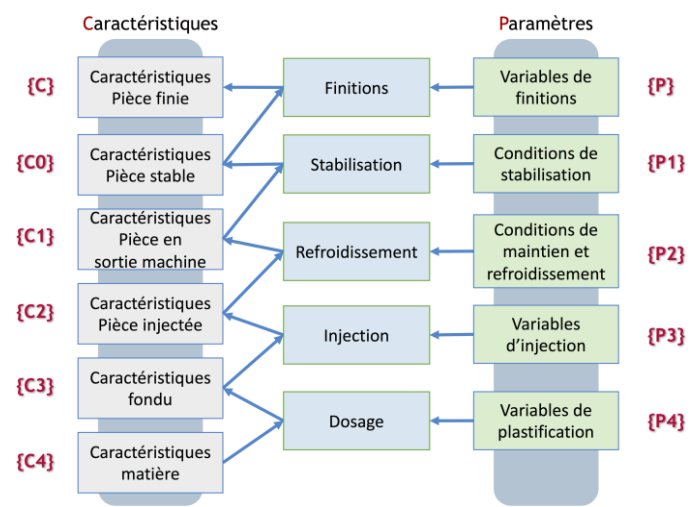

Figure 1. REPRESENTATION ZIG ZAG

## II. PILOTAGE PAR RESEAU DE NEURONES – ETAT DE L'ART

Afin de diminuer la dispersion de la production, les machines d'injection plastique sont mises sous contrôle automatique. Nous distinguons les méthodes de régulations des paramètres, procédés qui visent à garantir un point de fonctionnement et les méthodes d'ajustement des paramètres, dont le but est de trouver un point de fonctionnement optimal aux vues des caractéristiques qualités produites et des perturbations. Les réseaux de neurones permettent de modéliser des systèmes non linéaires aux multiples entrées et sorties avec des taux d'erreurs très faibles. Le procédé d'injection plastique possède des paramètres liés et non linéaires. Un réseau de neurones est un modèle relationnel composé de nœuds interconnectés (neurones), liés et pondérés. Le réseau le plus utilisé en injection des plastiques est le réseau à réaction par propagation inverse (feed-forward Back-Propagation Neural Network, BPNN) défini par Rumelhart [2]). Celui-ci se construit par apprentissage itératif : pour une entrée, la sortie donnée par le réseau est comparée à la sortie attendue. La différence obtenue est alors propagée depuis les nœuds de sortie vers l'entrée en ajustant successivement les coefficients de pondérations de chaque nœud. Le jeu de données de la phase d'apprentissage peut provenir de simulations ou de campagnes d'essais expérimentaux d'un outillage. Cette seconde proposition permet de prendre en compte des phénomènes qui ne sont pas encore modélisés dans les simulations. La réponse à une sollicitation sur un réseau entrainé est ensuite instantanée car ce n'est qu'un modèle relationnel.

### A. Régulation du procédé

Schnerr-Häselbarth et Michaeli [3] proposent le système *«Intelligent Quality Control»*. Il s'agit d'une interface informatique centralisant le pilotage et les mesures réalisées sur l'ensemble du parc machine et qui permet de réaliser les essais pour la phase d'apprentissage. Un algorithme est développé pour prédire la qualité des pièces produites ({C1} Figure 1) à partir des variables du procédé ({P2+P3+P4} Figure 1). Il s'agit ensuite de permettre l'exploitation des mesures pour réguler cycle à cycle. Les réseaux récurrents (*Recurrent Neural Networks*) de Jordan[4] et Elman[5] sont choisis car ils prennent en compte les corrélations temporelles. Des neurones spécialisés forment la couche contextuelle du réseau. Ces derniers ont la particularité de prendre en entrée leur propre sortie, en plus de l'entrée courante de connexion avec le reste du réseau ; ils réagissent ainsi en utilisant la mémoire de leur précédente valeur. Leur sortie dépend des entrées courantes et des entrées précédentes. L'apprentissage du réseau est réalisé en deux phases. La phase de conditionnement : les données sont entrées et le réseau propose une valeur en sortie. Celle-ci est ignorée. C'est la « couche contextuelle » du réseau JE qui mémorise et pondère les valeurs du réseau. Dans la seconde phase de sortie, c'est la couche contextuelle qui est la source des valeurs du réseaux (l'entrée est nulle). Le réseau produit alors une sortie. Cette sortie est comparée à la sortie attendue et les termes du réseau sont pondérés par propagation inverse. Ces deux étapes sont répétées pour l'ensemble des données composant la base d'apprentissage. Les données d'apprentissage proviennent d'un plan factoriel à trois niveaux faisant varier trois paramètres du procédé : température du fondu, pression de maintien et vitesse d'injection ({P4+P3+P2} Figure 1). 150 points de mesures de la pression dans le moule ({C2+C1} Figure 1) sont enregistrés pendant les phases d'injection et de maintien. La masse de la pièce est mesurée avec une précision de 1 milligramme et la plage d'essai couvre une variation de masse de 1,4%. Le réseau est alors capable de prédire la masse des pièces avec une exactitude de 86 à 95,2%, soit une moyenne de 7 milligramme d'erreur sur la masse. Les auteurs retiennent une topologie de réseau 1-10-10-1. Ils concluent leur étude en constatant que la topologie du réseau de neurone choisie a une influence minime sur le succès de la phase d'apprentissage. Avec deux couches de neurones cachées, plus le modèle possède de neurones, plus le modèle est correct. De plus l'organisation des nœuds cachés n'est pas importante. En revanche, les auteurs observent que des réseaux possédant trois couches cachées donnent des résultats légèrement moins justes que les réseaux à deux couches, sans proposer de comparaison numérique. Les auteurs expliquent ce phénomène par un nombre trop élevé de neurones. Par la suite, Michaeli et Schreiber [6] régulent la pression dans le moule en pilotant la pression hydraulique d'injection, à l'aide d'un réseau BPNN entrainé sur 15 cycles d'injection du système avec différents réglages. La température de la matière dans le moule est mesurée par capteur infrarouge ({C2} Figure 1) et la pression d'injection ({P3} Figure 1) est régulée afin de suivre le diagramme Pression Volume Température de transformation du matériau pendant l'ensemble du cycle. Les résultats expérimentaux obtenus montrent qu'une augmentation de 20 degrés de la température cause une augmentation de 0,07% de la masse de la pièce, ce qui semble négligeable. En comparaison, la même augmentation produit une diminution de 1.27% sans la régulation proposée.

### B. Ajustement des paramètres du procédé

Les problèmes de faisabilité d'un mesurage des pièces en cycle ont été résolus par la prédiction de la qualité des produits à partir des variables intermédiaires mesurée sur le cycle du procédé. Une mesure directe reste préférable pour plus de précision (*voir II.B.2*).

#### 1) Ajustement par les caractéristiques prédites

Haeussler et Wortberg [7] utilisent un réseau de neurones pour prédire la masse et la longueur des pièces produites à partir des paramètres du procédé. Ils entrainent un réseau 9-21-2 à partir des mesures sur une production de 162 cycles continus. Le réseau montre des résultats en corrélations avec la mesure de pièces de 80% pour la masse et 88,2% pour la longueur. Ces résultats sont comparés avec une régression non-linéaire qui produit des corrélations de 75% pour la masse et 79% pour la longueur. Ils concluent leur étude en identifiant la nécessité d'un réseau adaptatif pour un véritable contrôle en cycle industriel. Le réseau doit apprendre des mesures cycle à cycle et ne pas répondre qu'à des données apprises. Woll et Cooper [8] comparent ensuite les performances d'une analyse statistique MSP/SPC avec un réseau de neurones BPNN pour prédire la masse des pièces à partir de profils de paramètres discrets (pression hydraulique, dans le moule et dans la buse) et accepter ou non une pièce produite. Le réseau est entraîné comme un modèle inverse du procédé : des profils de pression de moule en entrée correspondent à des valeurs de paramètres de pression de maintien ou température du fourreau. Les paramètres sont alors ajustés cycle à cycle en comparant le profil de pression dans le moule actuel avec le profil de référence appris. L'analyse MSP est effectuée sur la valeur de pression maximale mesurée dans le moule tandis que le réseau de neurones est entraîné sur l'ensemble du profil de pression. Le réseau est entraîné sur des résultats simulés afin d'éviter de réaliser une longue campagne d'essai, la convergence du modèle étant observé pour 2000 essais. L'essai est réalisé pour un unique matériau et ne s'intéresse qu'à la longueur de l'éprouvette. Les résultats montrent la capacité supérieure des réseaux de neurones à prédire les non linéarités comparé à la régression linéaire multiple utilisée par SPC. L'exactitude de la prédiction des mesures est augmentée de 10% et le taux de faux positif est réduit de moitié. Le réseau répond très bien aux perturbations. L'étude conclue sur la nécessité d'entraîner le réseau sur plusieurs profils, dont les températures.

*2) Ajustement par les caractéristiques mesurées*

Il est possible de mesurer la qualité des pièces produites dès la sortie de moule. Dans ces études, la diversité des caractéristiques mesurées sur les pièces est limitée. La masse est la mesure récurrente car elle peut être réalisée pendant le cycle, et ainsi être utiliser pour corriger le cycle suivant. Les mesures complémentaires sont effectuées à posteriori de la production d'un lot, pour valider les performances du modèle de pilotage. Il serait intéressant de réaliser une caractérisation complète de la pièce obtenue pendant le temps de cycle afin de corriger les pièces suivantes à partir de ces données. Sous réserve d'une durée de mesurage des pièces inférieure au temps de cycle, le pilotage peut s'effectuer dès la pièce suivante. Lau et al. [9] proposent de vérifier les capacités de création par apprentissage d'un modèle multi entrées-sorties du procédé d'injection par réseau de neurones à propagation inverse. L'objectif est que le réseau suggère des modifications des paramètres procédés afin d'ajuster les dimensions de la pièce, pour obtenir la pièce cible en fonction de la pièce obtenue. L'étude dimensionnelle porte sur une longueur, deux largeurs et l'épaisseur d'un échantillon. L'apprentissage du réseau est réalisé sur les mesures de 100 pièces produites en faisant varier 6 paramètres procédés : la température du fourreau central et arrière, la vitesse de fermeture de l'outillage, la vitesse d'avance de la vis, la force de fermeture, la durée d'injection, la durée de refroidissement, la vitesse d'injection et la pression d'injection. L'étude souligne que malgré le maintien de paramètres de production identiques, les quatre dimensions étudiées varient au bout de quatre jours de productions, signe de la difficulté d'obtenir une production stable en injection et le besoin de contrôle. Les variations dimensionnelles mesurées sont appliquées en entrée au réseau, qui indique en sortie les valeurs du point de fonctionnement permettant de produire une pièce avec ces dimensions. Connaissant le réglage utilisé pour produire les pièces, l'étude propose de régler les paramètres dans des proportions inverses afin de compenser les variations de pièces. Cette démarche suppose que les paramètres soient indépendants, ce qui n'est pas le cas. Néanmoins, après 3 itérations de réglages successifs, le réseau parvient à converger vers une erreur quadratique moyenne (RMS) de 0,07. Ce résultat est comparé à une précédente étude des auteurs [10] qui utilisaient un système hybride à logique floue qui convergeait en 3 itérations vers une erreur quadratique moyenne (RMS) de 0,06. La méthode proposée est moins performante qu'un système à logique floue et réseaux de neurones mais plus simple à mettre en œuvre. De plus, une optimisation de la topologie du réseau de neurones, telle que la recherche en informatique la permet aujourd'hui améliorerait ce résultat (*voir III.B*).

## III. VERS UN PILOTAGE PAR LA QUALITE PRODUITE

Nous définissons la caractéristique qualité intermédiaire comme un critère qualité immédiatement mesurable en sortie de presse ({C1} Figure 1), non fonctionnel, mais fortement corrélé avec les caractéristiques fonctionnelles de la pièce finie. Le projet SAPRISTI vise à évaluer l'intérêt de l'exploitation de ces caractéristiques qualités intermédiaires. Nous proposons d'identifier les caractéristiques intermédiaires mesurables sur le procédé. Nous validerons ensuite la corrélation de ces caractéristiques avec la qualité de la pièce finie. Sur la base de ces travaux, nous proposons de développer un modèle du procédé prenant en compte l'ensemble des caractéristiques intermédiaires mesurables afin d'ajuster les paramètres du procédé. Ce modèle pourra être construit par apprentissage à l'aide de réseaux de neurones. Nous distinguons les caractéristiques {Ci} et les paramètres réglables {Pi} du procédé pour chacune des phases présentées dans la Figure 1. Notre étude nous permet d'identifier les caractéristiques intermédiaires du procédé régulièrement retenues dans la littérature (Tableau 1).

| Caractéristiques produits | | Paramètres du procédé | |
|---|---|---|---|
| *C* | Pièce finie (masse, géométrie) | | |
| *C1* | Caractéristiques intermédiaires | | *P1* |
| *C2* | Pression dans moule | Pression de maintien | *P2* |
| | Température dans le moule | | |
| *C3* | Température du fondu | Pression d'injection | *P3* |

Tableau 1. VARIABLES ET PARAMETRES DU PROCEDE

### A. Identification des paramètres influents

Afin de caractériser l'influence des paramètres réglables sur les caractéristiques des produits, nous réaliserons un plan d'expérience de Plackett-Burman [11] à deux niveaux. Ce plan nous permet d'identifier les paramètres influents par analyse statistique, par exemple en utilisant un test de Fisher. Nous mesurons le maximum de caractéristiques possibles, afin d'utiliser ces données pour l'apprentissage d'un modèle réseau de neurones. Nous retenons des facteurs procédés supposés pertinents afin de les faire varier. L'objectif de cette phase n'est pas de réguler le procédé autour d'un point de fonctionnement optimal mais de faire varier l'ensemble des paramètres afin de créer de la variabilité pour l'analyse statistique. Cette variabilité, permettra d'identifier et de hiérarchiser les caractéristiques pertinentes, mais aussi de créer de la variabilité sur les caractéristiques élémentaires … et sur les pièces finies. Nous utiliserons ainsi cette variabilité pour valider l'hypothèse de l'intérêt des caractéristiques intermédiaires comme prédicteurs de la qualité finie et comme intermédiaire de pilotage court terme du processus.

*B. Modélisation par apprentissage*

*1) Topologie du réseau de neurones*

Afin de définir la topologie du réseau de neurone employé dans notre étude, nous évaluerons différents nombres de couches cachées. Un nombre de couche trop élevé entraine un phénomène de sur-ajustement (*overfitting*) qui génère des erreurs dans la réponse alors qu'un nombre trop faible ne permet pas de modéliser précisément le problème. Après avoir entraîné différents réseaux sur nos données, nous choisissons la configuration qui minimise l'erreur quadratique moyenne lors de l'apprentissage mais aussi lors de l'évaluation du réseau sur des données expérimentales. Rivals [12] propose d'optimiser ce nombre en deux phases. La première dans laquelle le nombre de couches cachés est augmenté jusqu'à observer le sur-ajustement, puis une seconde phase dans laquelle les réseaux retenus sont évalués par test de Fisher pour déterminer si l'ensemble des nœuds cachés sont statistiquement pertinents. Cette méthode permet une optimisation rapide de grands réseaux. Xu et Chen [13] évaluent différentes méthodes d'optimisation du nombre de couches cachées d'un réseau de neurones et proposent une approche reprise dans de nombreuses publications. Néanmoins, la méthode proposée ne prend pas en compte la quantité de bruits des données, le type de fonction d'activation ou l'algorithme d'entrainement utilisé.

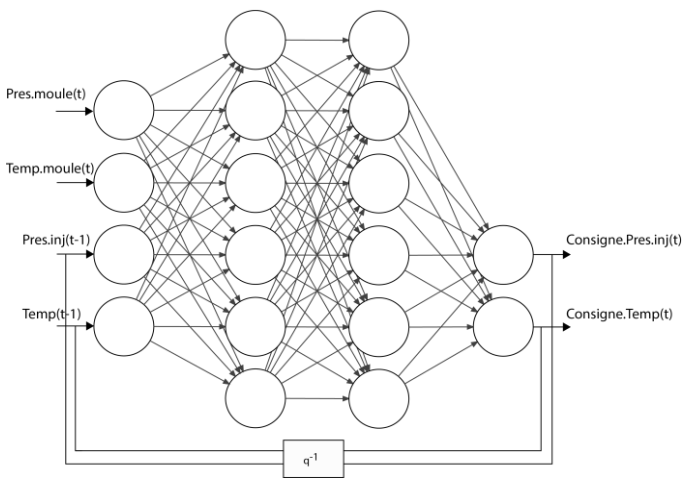

Figure 2. RESEAU 4-6-6-2 DYNAMIQUE

*2) Réseau de neurones dynamiques*

Nous proposons d'évaluer l'utilisation d'un réseau de neurone récurrent (bouclé) prenant en compte la variable temporelle, afin d'exploiter les mesures discrètes pendant les cycles (température, pression, déplacement). Ces réseaux demandent de postuler l'influence des perturbations (bruit) sur le procédé. Dans notre cas, le bruit est additif avec la sortie actuelle et l'ensemble des sorties passées (hypothèse NARX : Nonlinéaire AutoRégressif à entrées eXogènes). Pour illustrer la construction d'un réseau dynamique, la Figure 2 représente un réseau dynamique 4-6-6-2 capable de déterminer des consignes de pression et température à partir des mesures obtenues pendant le cycle dans le moule. L'échantillonnage des mesures réalisées dans le moule pendant le cycle est proche de 100 mesures par secondes, le réseau devra être dimensionné en conséquence et les mesures moyennées par intervalles. Nous utilisons des capteurs de pression capacitifs et des thermocouples. Les capteurs de pression ont une dérive inférieure à 1 bar par secondes ce qui est satisfaisant sur des temps de cycle inférieurs à 30 secondes.

*C. Mesurage de la qualité des pièces en cycle*

Le procédé d'injection est cyclique (Figure 3). Les différentes phases du procédé, comme le dosage de la matière pour la prochaine pièce et le refroidissement de la pièce actuelle sont réalisés en temps parallèles. Afin de respecter le temps de cycle, le mesurage de la pièce produite doit être réalisé entre l'éjection de la pièce de l'outillage et l'éjection de la pièce suivante. La caractéristique des pièces finies la plus étudiée dans la littérature est la masse. Cela s'explique par la facilité de mesure en sortie machine et la bonne précision des balances (milligramme). La masse est une condition nécessaire pour valider les pièces. Les caractéristiques d'aspect sont en revanche peu étudiées. Elles contiennent des informations capitales pour l'analyse des causes de la variabilité d'une pièce et intéressent le client. Le régleur utilise prioritairement l'aspect des pièces pour régler.

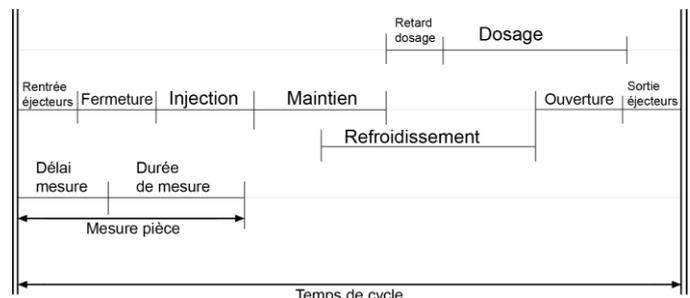

Figure 3. CHRONOGRAMME DU CYCLE DU PROCEDE

Le contrôle qualité géométrique par imagerie est appliqué pour les grandes séries. Il demande un équipement calibré et automatisé. Il prend place après la production, permet d'exclure les pièces non conformes et ne s'intéresse pas à l'aspect. Les dimensions mesurées ne sont pas utilisées pour ajuster le procédé pour les cycles suivants. En parallèle, le manque de spécification et la difficulté à mesurer, en cycle, limite l'utilisation de ces caractéristiques. De plus, aucune norme ne statue actuellement sur les qualités d'aspect. Plusieurs travaux de recherches sont en cours [14], [15]. Ces travaux associent une démarche de spécification aux développements de solutions de mesures adaptées aux cycles industriels [16]. Différentes mesures sont aujourd'hui compatibles, de par leurs courtes durées, avec une utilisation en cycle industriel : masse, thermographie infrarouge, contrôle d'aspect, imagerie tridimensionnelle, durométrie. Pour obtenir des mesures fiables et répétables cette opération doit être automatisée. En sortie de moule, la pièce est chaude et déformable. De nombreuses presses industrielles sont équipées de robot de déchargement. Nous proposons de programmer la cinématique du robot afin de déplacer la pièce devant différents outils de mesures (Figure 4). Ainsi nous synchronisons le mesurage de toutes les pièces en sortie de moule à des instants fixes. Nous obtiendrons des mesures répétables sur l'ensemble des cycles d'injection. Nous utilisons un outil de mesure laser par nuage de point d'une résolution de 0,050 millimètres entre points. Nous pourrons ainsi caractériser une variation dimensionnelle maximale de 1,5% sur nos échantillons. En comparaison, la norme AFNOR NF T58-000 [17], pour les dimensions de nos pièces, définie une tolérance dimensionnelle très large de 13% en classe *précision*.

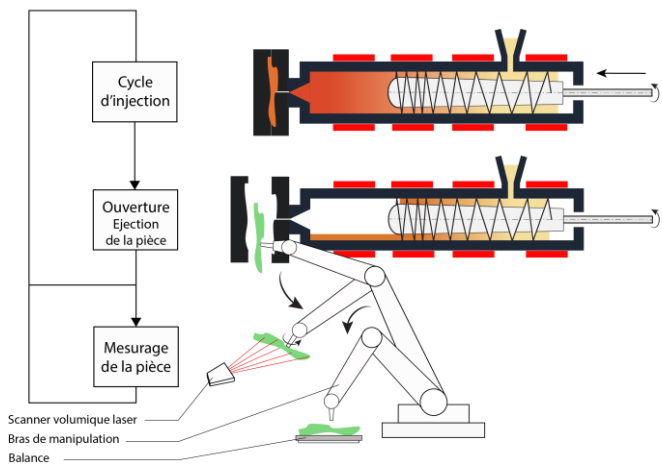

Figure 4.   MESURAGE DES PIECES EN CYCLE

Nous réaliserons cette mesure dimensionnelle dès la sortie de moule. Elle sera ensuite réalisée après stabilisation de la pièce, une heure puis une journée après la production, afin d'analyser les phénomènes de retraits. Nous compléterons nos mesures par imagerie thermographique infrarouge en sortie de moule des pièces afin d'analyser les champs de températures qui pourraient être corrélés à la qualité. Une mesure de la qualité d'aspect dès la sortie de moule est envisageable par prises de vues photographiques et analyse d'images. L'aspect de surface des pièces sera mesuré et caractérisé post-production sur un dôme de mesure de réflectance (*Polynomial Texture Mappings*) [18]. L'ensemble de ces mesures nous permettra d'identifier des corrélations entre caractéristiques et paramètres, ainsi que d'entrainer le réseau de neurones du modèle.

IV. CONCLUSION

Le pilotage du procédé d'injection est une thématique de recherche riche et aux enjeux industriels importants. Les méthodes de pilotage s'appuient sur l'instrumentation de la machine et nous proposons d'automatiser le mesurage des pièces produites. L'objectif est d'obtenir un ajustement cycle à cycle des paramètres procédé afin d'obtenir un produit à la qualité cible. Le procédé d'injection possède des paramètres dont les effets sur la géométrie sont non-linéaires. Les réseaux de neurones permettent de modéliser par apprentissage un procédé multi entrées et sorties, à partir de données expérimentales. La littérature en pilotage du procédé d'injection plastique propose des études précises sur des phases uniques du processus et ne prend pas en compte l'ensemble des relations entre les phases (Figure 1), ni temporelles. Le lien entre les paramètres mesurables sur la presse et la qualité finale des produits est peu étudiée. Enfin, les critères sensoriels des produits ne sont pas pris en compte alors qu'ils sont essentiels pour la qualité des produits. La recherche en métrologie géométrique et d'aspect automatisée et la recherche active dans le domaine des réseaux neuronaux ouvrent de larges perspectives pour le pilotage. L'apport de la modélisation dynamique par réseaux de neurones permet une réponse très rapide rendant compatible le pilotage cycle à cycle.